\documentclass{elsart}

\usepackage[dvips]{graphics}
\usepackage{amssymb}

\newcommand*{\difd}{\d}                      
\newcommand*{\figur}{Fig.\nolinebreak{\hspace{.3em}}} 
\newcommand*{\anhang}{Appendix\nolinebreak{\hspace{.3em}}} 
\newcommand*{\gleichung}[1]{Eq.\nolinebreak{\hspace{.3em}}(#1)} 
\newcommand*{\gleichungen}[1]{Eqs.\nolinebreak{\hspace{.3em}}(#1)}
\newcommand*{\tabelle}{Table\nolinebreak{\hspace{.3em}}} 
\newcommand*{\nummer}[1]{Nr.\nolinebreak{\hspace{.3em}}#1} 
\newcommand*{\pictop}{\textbf{(top)} }       
\newcommand*{\picbottom}{\textbf{(bottom)} } 
\newcommand*{\picleft}{\textbf{(left)} }     
\newcommand*{\picright}{\textbf{(right)} }   
\newcommand*{\pica}{\textbf{(a)} }           
\newcommand*{\picb}{\textbf{(b)} }           
\newcommand*{\picc}{\textbf{(c)} }           
\newcommand*{\picd}{\textbf{(d)} }           

\journal{Powder Technology}

\begin{document}

\begin{frontmatter}
\title{Statistical model of the powder flow regulation by
	nanomaterials}

\author[physik]{D. Kurfe\ss\corauthref{cor}},
	\ead{physik@kurfess.net}
\author[physik]{H. Hinrichsen},
\author[pharmazie]{I. Zimmermann}

\address[physik]{Fakult\"at f\"ur Physik und Astronomie,
	Universit\"at W\"urzburg, Am Hubland, D-97074 W\"urzburg,
	Germany}
\address[pharmazie]{Fakult\"at f\"ur Chemie und Pharmazie,
	Universit\"at W\"urzburg, Am Hubland, D-97074 W\"urzburg,
	Germany}

\corauth[cor]{Corresponding author.}

\begin{abstract}
Fine powders often tend to agglomerate due to van der Waals forces
between the particles. These forces can be reduced significantly by
covering the particles with nanoscaled adsorbates, as shown by recent
experiments. In the present work a quantitative statistical analysis of
the effect of powder flow regulating nanomaterials on the adhesive
forces in powders is given. Covering two spherical powder particles
randomly with nanoadsorbates we compute the decrease of the mutual van
der Waals force. The dependence of the force on the relative surface
coverage obeys a scaling form which is independent of the used
materials. The predictions by our simulations are compared to the
experimental results.
\end{abstract}
\end{frontmatter}


\section{Introduction}

Dry granular powders are basic materials of great importance to the
pharmaceutical industry. {\em Bulk\/} powders tend to agglomerate
due to adhesive forces between the powder particles. This
characteristic causes problems in the manufacturing process of
drugs, where accurate dosing is essential \cite{meyer_zimmermann}.
Furthermore, insufficient flowability can severely harm production
devices as well as prevent inhalable drugs from reaching the lung's
alveoli.

\begin{figure*}
	\includegraphics*{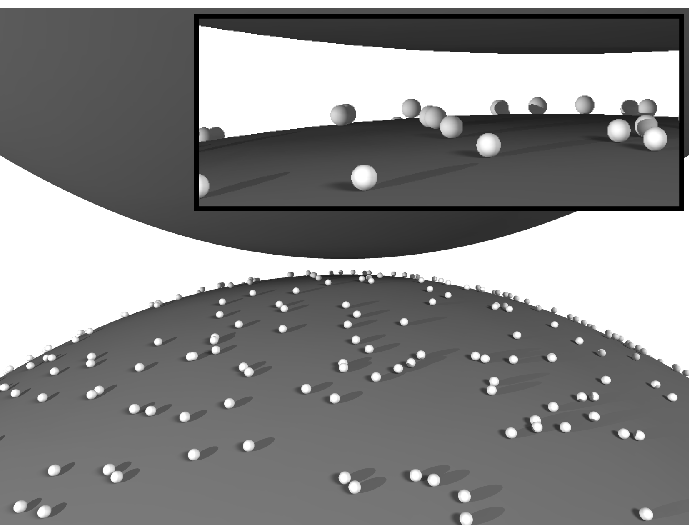}
	\includegraphics*{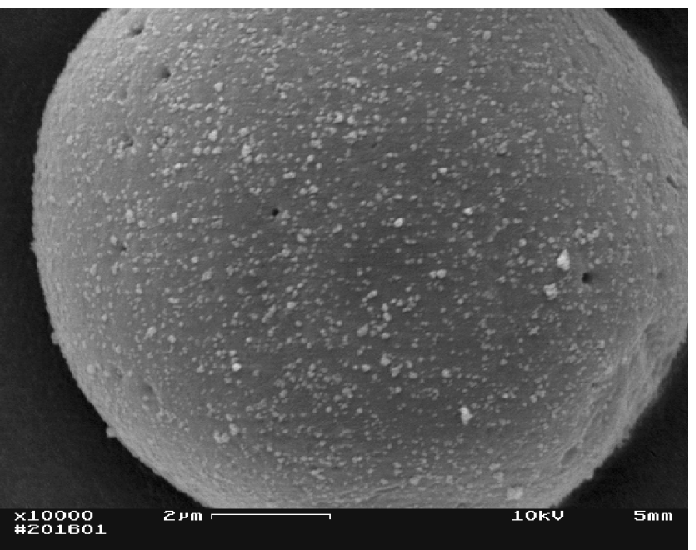}
\caption{\picleft Model of a covered powder particle of radius
$R=11\,\mathrm{\mu m}$ and nanoadsorbates of $r=50\,\mathrm{nm}$
(relative surface coverage of $2\%$ on the lower sphere). Above, a
second uncovered particle is approaching. The inset displays a
zoomed view of the contact field. \picright Microscopic photo of a
corn starch particle covered with aggregates of the nanomaterial
Aerosil\textsuperscript{\textregistered} 300, taken from
\cite{diss_meyer}.
}
\label{figCoatedPowderParticle}
\end{figure*}

Pharmaceutical powders typically have particle diameters on the
micrometer scale and contact distances between the particles on the
angstrom scale. If they are dry and not highly electrified, van der
Waals forces are the strongest interparticle forces, exceeding
gravitational and Coulomb forces significantly. As van der Waals
forces are short-range it is possible to effectively decrease these
adhesive forces by increasing the surface roughnesses of the powder
particles by covering them with nanoparticles (cf.\ 
\figur\ref{figCoatedPowderParticle}). This was verified by tensile
strength experiments, in which two powder layers were separated and
the required force was measured \cite{diss_meyer,zimmermann_et_al}.

H.C. Hamaker calculated the van der Waals forces between
macroscopic spherical bodies by integration over all molecular
dipole-dipole interactions \cite{hamaker}. Accordingly
\begin{equation}
	F_{\mathrm{vdW}} = - \frac{A}{6 H^2} \frac{R_1 R_2}{R_1+R_2}
\label{eqHamaker}
\end{equation}
is the attractive interparticle force between two spheres of radii
$R_1$ and $R_2$, where all material constants are combined into the
so called Hamaker constant $A$. $H$ is the distance between the
surfaces of the two bodies. In order to avoid a diverging force upon
contact, one usually assumes that $H$ cannot become smaller than a
so called contact distance $D=4\times 10^{-10}\,\mathrm{m}$
\cite{rumpf,israelachvili,zimmermann_et_al}.

K. Meyer theoretically described the decrease of the tensile
strength with increasing surface coverage assuming a stable
three-point contact, i.e., effectively three nanoparticles act as
spacers between two powder particles and combine them
\cite{diss_meyer}. In bulk powders, two particles firstly joined
by a one- or two-point contact are subject to torque, and their
bonding remains unstable until a three- (or more-)point contact
is reached.

Meyer considered an idealized system in which the three contact
adsorbates are fixed on the vertices of an equilateral triangle
\cite{diss_meyer,zimmermann_et_al}. This model allows one to
calculate analytically the distance of the two bulk powder particles
against the side length of this triangle. From this distance the
interparticle force can immediately be derived using Hamaker's
sphere-sphere model.

The result of this calculation describes only two states sharply
separated by a critical surface coverage $\rho_c$. For
$\rho<\rho_c$ the powder particles touch and attract each other
strongly, while otherwise the nanomaterial establishes a gap
between them and the short-range adhesive force vanishes nearly
entirely. Obviously Meyer's model of equidistant spacing of the
nanoparticles ignores any stochastic aspects which are relevant in
reality.

It is the aim of this paper to give a more realistic theoretical
description of the effect of flow regulating nanomaterials on the
interparticle forces acting between the larger powder particles. We
simulate {\em random\/} coverages of the surfaces of the host
particles with nanoparticles and calculate the van der Waals forces
averaged over many realizations of randomness. We use spherical host
particles and nanoadsorbates as a first approximation of reality, as
suggested by microscopic photos of corn starch powder particles
mixed with nanomaterials, see \figur\ref{figCoatedPowderParticle}.
Our generated data indicates a {\em scale invariant\/} behavior,
where an incomplete gamma function fits the resulting scaling
function perfectly. Our results are compared to experimental data,
demonstrating the improvements compared to Meyer's model.

\section{Simulation}

The simulation computes the averaged van der Waals force
$F_{\mathrm{vdW}}$ versus the relative surface coverage $\rho$. The
algorithm consists of three parts: (i) the preparation of a host
particle by covering it randomly with adsorbates, (ii) the
positioning of two host particles so that a stable three-point
contact is formed, and (iii) the computation of the van der Waals
force by summation over all resulting sphere-sphere interactions
according to the Hamaker model. So each run of this algorithm gives
one sample of the adhesive force which can occur.

In the following the three steps of the simulation are shortly
explained (for more detailed information see
\anhang\ref{appSimulation}).

\subsection{Preparation} Initially, a host sphere is randomly and
successively covered with nanoparticles, so that a uniform
(Poisson) distribution with respect to the spherical surface area
is established. Unlike the random sequential adsorption (RSA)
model, our model does allow the adsorbate particles to overlap. Yet
for the low coverages which we are interested in ($\rho$ has the
order of magnitude of few percent), the probability of two
overlapping nanoparticles is negligible.

\subsection{Positioning} 
\begin{figure*}
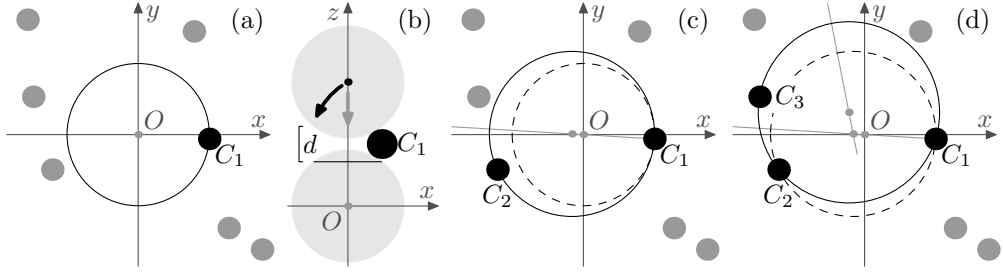

	\includegraphics*{graphics/posit1.ps}
	\includegraphics*{graphics/posit2.ps}
	\includegraphics*{graphics/posit3.ps}
	\includegraphics*{graphics/posit4.ps}
\caption{Algorithm for finding the three contact points. \pica Top
view on the covered host particle (white background) and \picb side
view. The large dots in (a) are randomly distributed adsorbate
particles. The uncovered powder (host) particle is approaching from
above along the $z$-axis. The first adsorbate to be hit by the
approaching sphere is the one which is located nearest to the
$z$-axis ($C_1$). Upon contact the circle in (a) describes the line
of equal distance $d$ in $z$-direction between the two host
particles. $C_1$ withstands the approaching sphere, so that the
momentum of this moving sphere results in a torque (black arrow).
\picc Due to the torque the upper host particle rolls off so that
its center projected onto the $x$-$y$-plane moves along
$\overline{C_1O}$, until this host sphere hits the second contact
adsorbate ($C_2$). \picd Now $C_1$ as well as $C_2$ hold the
approaching host sphere so that its center (projected onto the
$x$-$y$-plane) moves along the perpendicular bisector of the side
$C_1C_2$, until a stable three-point contact is formed.
}
\label{figPositioning}
\end{figure*}

In the second part of the algorithm a
covered host and (for simplicity reasons) an uncovered powder sphere
are brought to collide, cf.\ \figur\ref{figCoatedPowderParticle}
left. The uncovered large powder sphere rolls off over the small
adsorbates, until a stable three-point contact builds up (see
\figur\ref{figPositioning}); or, if the three potential contact
candidates are too far away from each other, it rolls off until the
two powder spheres directly touch each other. Knowing the
characteristics of the contact, it is easy to calculate the new
position $\vec{s}$ of the center of the uncovered powder sphere
relative to the center of the covered host sphere, which is assumed
to be located at the origin.

\subsection{Computation of the force} The coordinates $\vec{a}_i$
describe the center positions of the $N_A$ adsorbates on the covered
host particle. Let $R$ be the radius of the powder particles, and
$r$ the common radius of all nanoadsorbates. $A_{HH}$ and $A_{AH}$
are the host-host and the adsorbate-host Hamaker constants,
respectively. By vectorial summation over all sphere-sphere
interactions according to \gleichung{\ref{eqHamaker}} with two
(host) powder particles, we get
\begin{equation}
	\vec{F}_{\mathrm{vdW}} = - \frac{A_{HH}}{6 \,
			(|\vec{s}|-2 R)^2} \frac{R}{2} \cdot
			\frac{\vec{s}}{|\vec{s}|}
			- \sum^{N_A}_{i=1} \frac{A_{AH}}{6 \,
			(|\vec{s}-\vec{a}_i|-r-R)^2} \cdot
			\frac{\vec{s}-\vec{a}_i}{|\vec{s}-\vec{a}_i|}
\end{equation}
as the resulting interparticle force.

Note that not only the three contact adsorbates but also all other
adsorbate particles without direct contact to the uncovered powder
sphere are taken into account. Most of them give only minor
contributions to the adhesive force, as they are far away from the
three-point contact and are therefore distant from the uncovered powder
sphere. However, statistically some of them may come very close to the
three-point contact and give force contributions of significant order of
magnitude. Thus for a complete treatment of randomness it is reasonable
to consider the influence of all the adsorbate particles for the
computation of the force.

\section{Results and discussion}

For the further discussion, $A=A_{HH}=A_{AH}$ is assumed; most
condensed phases, solid as well as liquid ones, have Hamaker
constants of about $10^{-19}\,\mathrm{J}$ interacting across vacuum,
and similar values in air
\cite{schubert,israelachvili,zimmermann_et_al}.

\begin{figure*}
	\includegraphics*{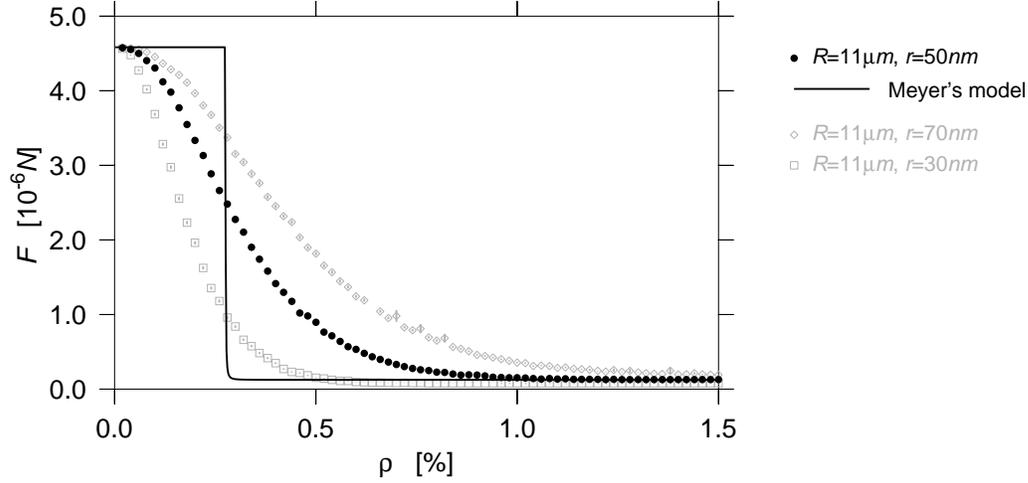}
\caption{Simulated data of the van der Waals force versus the
relative surface coverage ($A=8\times 10^{-19}\,\mathrm{J}$, mean of
$10^4$ samples per point and standard error plotted). For comparison
Meyer's model for $R=11\,\mathrm{\mu m}$ and $r=50\,\mathrm{nm}$ is
displayed.
}
\label{figData1}
\end{figure*}

Concerning the adhesive effect between two particles, we have to
examine only the absolute value of the van der Waals force,
$F=\langle|\vec{F}_{\mathrm{vdW}}|\rangle$ as the mean of many
samples generated by our algorithm.  Some results of our simulations
are shown in \figur\ref{figData1} (for details about the comparison
with Meyer's model cf.\ \anhang\ref{appMeyer}).

\subsection{Scaling}

	\subsubsection{Scaling properties}

For $\rho\rightarrow 0$, the force approaches the adhesion between
two uncovered powder spheres,
\begin{equation}
	F \rightarrow \, F_0 = \frac{A \cdot R}{12 \, D^2}
	\, \propto R \quad .
\end{equation}
In \figur\ref{figData1} a linear right shift of the decrease of $F$
in dependence of the surface coverage $\rho$ for increasing $r$ can
be observed. This suggests the scaling form
\begin{equation}
	\frac{F}{R} \approx f\left(\frac{\rho R}{r}\right) \quad .
\end{equation}

\begin{figure*}
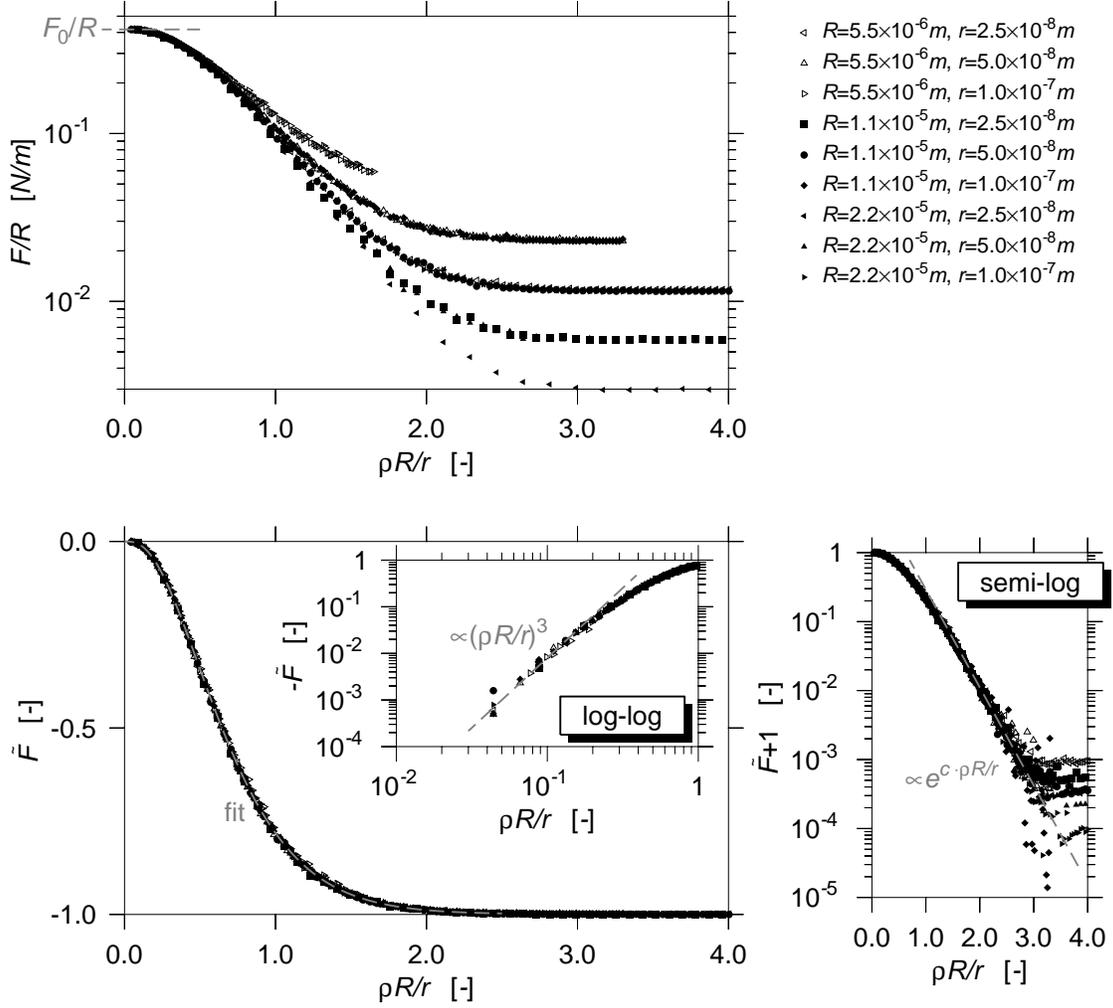

	\includegraphics*{graphics/data2.ps}
	\\[5mm]
	\includegraphics*{graphics/data2uni.ps}
\caption{\pictop Semi-logarithmic plot of the rescaled simulated
data ($A=8\times 10^{-19}\,\mathrm{J}$, mean of $10^4$ samples per
point). Increasing $\rho R/r$ the curves saturate at a constant
which scales approximately as $\propto r/R$. The plot indicates a
general behavior for $\rho R/r\rightarrow 0$. \picbottom The
converted quantity $\tilde{F}=\frac{F-F_0}{F_0-F_m}$ of the same
data produces a convincing data collapse. The inset and the
right-hand plot indicate a power law behavior for very small
$\rho R/r$ and an exponential decay shortly before the minimum is
reached, respectively. The scaling function is fitted by
$-0.499692\cdot (2-\Gamma(3, \, 4.144 \, \rho R/r))$.
}
\label{figData2}
\end{figure*}

The collapse of the data in \figur\ref{figData2} (top) shows that
this scaling form works well for small values of $\rho R/r$, while
for small $F/R$ the curves become constant at different values,
which grow roughly $\propto r/R$. This seems plausible, because with
a large gap between the two host particles the force is mostly
affected by the three contact adsorbates only:
\begin{equation}
	\min (F) \approx F_m = 3\cdot \frac{A}{6 D^2} \frac{R r}{R+r}
	\, \propto r
\end{equation}
for $r\ll R$.

Our previous thoughts call for an improved ansatz of the form
\begin{equation}
	\tilde{F} = - \, \frac{F_0(R)-F(\rho,R,r)}{F_0(R)-F_m(r)} =
	g\left(\frac{\rho R}{r}\right) \quad ,
\end{equation}
where $g$ is a scaling function in the range $0\le\rho R/r\lesssim 2.5$.
Here the difference of the offset $F_0$ to the data $F$ is determined
and divided by the possible maximum of this difference, given by
$F_0-F_m$. Scaling this way would give positive values with a maximum of
$1$. In order to obtain a function whose curve represents the decrease
of the adhesive force with increasing $\rho$ in a more intuitive manner,
the minus sign is inserted. Now, as can be seen in \figur\ref{figData2}
(bottom), we obtain a perfect collapse of the data. Note that
$\tilde{F}$ does not depend on $A$, i.e., it is independent of the used
materials.

	\subsubsection{Asymptotic properties of the scaling function}

For $\rho R/r\rightarrow 0$ we expect a decrease of $\tilde{F}$
which is $\propto\rho^3$, as can be explained as follows. In the
limit of {\em low densities\/}, $\rho\rightarrow 0$, a decrease of
the adhesive force can only occur if three potential contact
adsorbates lie within a critical area between the two
approaching host spheres. Assuming the positions of the adsorbate
particles to be uncorrelated, the probability of this event grows
$\propto\rho^3$. A logarithmic plot verifies this power law
behavior, cf.\ the inset of \figur\ref{figData2} (bottom).

Increasing $\rho R/r$ the semi-logarithmic plot in the right panel
of \figur\ref{figData2} indicates a usual exponential decay until
the minimum in $\tilde{F}$ is reached at $\rho R/r\approx 2.5$.

	\subsubsection{Fit of the scaling function}

In our simulations the nanoscaled adsorbates are distributed
randomly on the spherical host particle, i.e., they satisfy Poisson
distributions in two dimensions, projected onto a spherical
geometry. In the analysis of 2D Poisson Voronoi cells it has been
demonstrated that the distribution of cell areas can be fitted to
gamma distribution functions with one, two or three parameters
\cite{weaire_et_al,kumar_kurtz,hinde_miles,marthinsen}. This
suggests an ansatz of the form
\begin{equation}
	g'\left(\frac{\rho R}{r}\right) =
	\alpha \, \left(\frac{\rho R}{r}\right)^{\beta}
	\exp\left(-\gamma \, \frac{\rho R}{r}\right)
\end{equation}
for the derivative of the scaling function $g$, where
$\alpha,\beta$ and $\gamma$ are constants. Integration yields
\begin{eqnarray}
	g\left(\frac{\rho R}{r}\right)
	&=& \int^{\frac{\rho R}{r}}_0 \alpha \, x^{\beta}
		\exp(-\gamma x) \, \difd x		\nonumber\\
	&=& \alpha \, \gamma^{-\beta-1} \Biggl(\Gamma(\beta+1) -
		\Gamma\left(\beta+1, \, \gamma \,
		\frac{\rho R}{r}\right)\Biggr) \quad .
\label{eqModelG}
\end{eqnarray}
Here $\Gamma(z)$ is the (complete) gamma function, and
$\Gamma(a, \, z)$ is the (upper) incomplete gamma function.

Since $\tilde{F}$ decreases as $\rho^3$ if $\rho$ is small (see
above), we expect the exponent $\beta=2$. Indeed fitting
\gleichung{\ref{eqModelG}} to our data in the range
$0\le\rho R/r\le2.5$, one obtains $\beta=2.00\pm 0.01$. Setting
$\beta=2$ and performing another fit, we determined the fit
parameters as $\alpha=-35.56\pm 0.07$ and $\gamma=4.144\pm 0.003$.
So the numerical result for the scaling function is
\begin{equation}
	g\left(\frac{\rho R}{r}\right) =
		-0.500\cdot \Biggl(2-\Gamma\left(3, \, 4.14 \,
		\frac{\rho R}{r}\right)\Biggr) \quad .
\end{equation}
In \figur\ref{figData2} (bottom) one can see that it matches the
rescaled numerical data perfectly.

\subsection{Comparison to experiments}

\begin{figure*}
	\includegraphics*{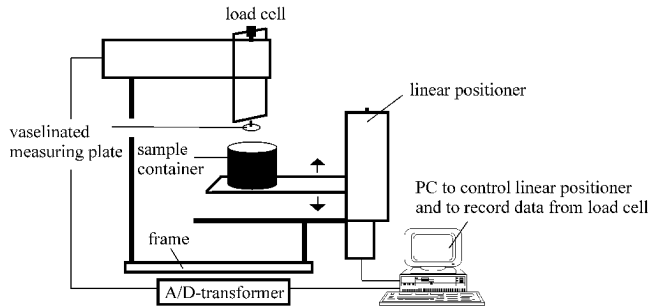}
\caption{Scheme of the tensile tester.
}
\label{figTensileTester}
\end{figure*}

The theoretical results can be compared to previous experiments
with a tensile tester \cite{diss_meyer}, cf.\ 
\figur\ref{figTensileTester}.

In this experiments a vaselinated measuring plate is brought into
contact with the surface of a powder sample. Lifting the plate a
tensile force is measured which allows one to determine the
interparticle forces needed to separate two powder layers (for
details see \anhang\ref{appExperiments}).

\begin{figure}
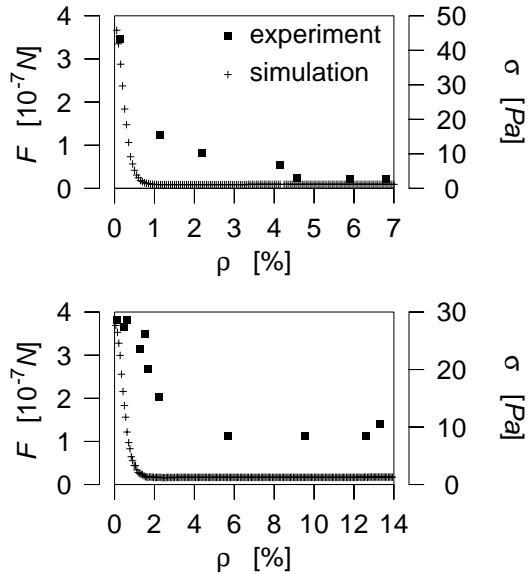

	\includegraphics*{graphics/aero300.ps}
	\\[2mm]
	\includegraphics*{graphics/aeroox50.ps}
\caption{Qualitative comparison of the simulated adhesive force $F$ with
experiments in which the tensile strength $\sigma$ was measured; note
that these are two different physical quantities $F$ and $\sigma$,
scaled against each other arbitrarily in order to give combinations
showing common trends in simulation and experiment. The experimental
data is taken from \cite{diss_eber}. \pictop After some mixing, by which
agglomerates of the nanomaterial are crushed into smaller aggregates,
aggregates of Aerosil\textsuperscript{\textregistered}~300 deposited on
corn starch powder particles ($R=11\,\mathrm{\mu m}$) have average radii
of $r=42\,\mathrm{nm}$. The simulation used $A=6.5\times
10^{-20}\,\mathrm{J}$ as Hamaker constant for this material combination.
\picbottom Analogous graph for
Aerosil\textsuperscript{\textregistered}~OX~50 on corn starch with
$R=11\,\mathrm{\mu m}$, $r=80\,\mathrm{nm}$ and $A=6.5\times
10^{-20}\,\mathrm{J}$.
}
\label{figAerosils}
\end{figure}

A comparison of the experimental data with the results of the simulation
is shown in \figur\ref{figAerosils}. Here the quantity of the simulated
adhesive force $F$ is arbitrarily scaled against the quantity of the
experimentally measured tensile strength $\sigma$ so that the offsets
collapse. For $\rho=0$ there are no nanoscaled adsorbates on the host
particle, and so $F$ and $\sigma$ are generated only by powder-powder
interactions. Although the experimentally measured forces for $\rho>0$
are systematically larger than the predicted ones, they show a
qualitatively similar type of curvature, confirming the approach of
studying random coverages.

At present the origin of the systematic deviations is not yet entirely
clear. They could be attributed to the influence of additional liquid
bridges between the powder particles. However, with corn starch the
formation of liquid bridges is only observed at relative humidities of
more than $40\%$ \cite{diss_eber}, and all experimental measurements
were performed well below this critical limit.

Aggregation is another important aspect. The adsorbate particles
mentioned above are aggregates of nanoscaled primary particles
\cite{diss_meyer}. Depending on the number of combined primary particles
the diameters of the adsorbates vary. The radii $r$ used in our
simulations are averages over many adsorbates whose sizes were measured
by atomic force microscopy during the experiments. Experimentally the
measured radii have errors of the order of about $r/2$ \cite{diss_eber}.
However, in the simulation statistical fluctuations of the adsorbate
radii are not included in order to keep the problem simple and fast
computable. In this way we make systematic errors, but a relative error
in $r$ of about $0.5$ alone cannot explain the quantity of the
deviations between simulation and experiment, cf.\
\figur\ref{figAerosils}.

The precise three-dimensional shapes of the adsorbates are also not yet
completely analyzed. Experimentally the aggregates are placed onto the
surfaces of the host particles by simply mixing the nanomaterial with
the powder. The number of adsorbed aggregates increases during mixing,
and the surface coverage of the host particles grows. As an unintended
side effect of this mixing a `grinding' of the aggregates occurs
\cite{meyer_zimmermann}. While at the beginning of the mixing process we
expect the shape of the adsorbed aggregates to be spheric as an
approximation, their shape gets continuously flatter due to the
mechanical stress during mixing. So we expect a deviation upwards from
the computed adhesive force because flat adsorbates generate a smaller
distance between two powder particles than the spherical adsorbates
assumed in the simulation.

Additionally, it seems to be likely that the experimental method
overestimates the actual forces. In fact, the top layer separated by the
tensile tester is not a monolayer of spheres. Instead we expect a rough
and irregular surface, especially at the edges of the sample, and so the
force needed to produce such a surface is presumably higher.

Nevertheless the systematic nature of the deviations allows us to
conclude that the simulation results may serve as {\em lower bounds\/}
to the actual forces and to the minimum surface coverage needed to
optimize the powder flow.

\section{Conclusions}

In the present work we have investigated how van der Waals forces
between two spherical particles are reduced by randomly deposited
nanoscaled adsorbates. In contrast to an earlier work by Meyer, who
considered a triangular coverage, the statistical analysis of
{\em randomly\/} deposited nanoadsorbates leads to an improved
prediction of the force reduction as a function of the coverage.
More specifically, a continuous curve instead of a step-like
function is obtained. Varying parameters we have identified scaling
laws which are independent of the used materials. A comparison with
previously recorded experimental data shows that the force predicted by
the simulation may be used as a lower bound for the actual force, though
the experimental data is not sufficient and not precise enough to verify
the functional behavior properly. This could be a starting point for
further experiments. Future simulations could include a statistical
distribution of the adsorbate size. However, the decline of the adhesive
force computed in our model may act as a lower approximation of the
minimum surface coverage which is needed to optimize the powder flow.

\appendix

\section{Simulation in detail}\label{appSimulation}

\subsection{Preparation}

For small $\rho$ there is no difference -- respecting the problem --
whether to regard two covered powder particles, or one uncovered
particle and one which is covered at double density, cf.\ 
\figur\ref{figCoatedPowderParticle}. Thus, for simplicity, the
preparation part of the simulation covers only one host particle
stochastically with the coverage density $2\rho$.

The center of the sphere to be covered is the origin in our
coordinate system. Random sphere point picking must be done with
respect to the sphere's surface area. In terms of spherical
coordinates, where $\theta$ is the polar and $\varphi$ the
azimuthal angle,
\begin{equation}
	\difd\Omega = \difd\varphi \, \difd\theta \sin\theta
		= - \difd\varphi \, \difd(\cos\theta)
\end{equation}
is an area element on the unit sphere. So a uniform point
distribution is achieved by rectangular distributions of
\begin{equation}
	\varphi \in [0,2\pi) \quad \mbox{and} \quad
	\cos\theta \in [-1,1] \quad .
\end{equation}

As a further simplification, we cover only one half of the first
sphere, where the contact to the second, uncovered sphere will be
made. The neglected adsorbates would hardly contribute to the
bonding force, as they would be too far away from the second powder
particle.

In this way adsorbates are successively placed upon the host, with
their centers at coordinates $\theta_i,\varphi_i$, until the
desired coverage is reached. Let $O_S=4\pi R^2$ be the host's
surface area and $C_A=\pi r^2$ the area of perpendicular projection
of an adsorbate on this surface. So the number of adsorbates on the
host sphere can be calculated from the relative surface coverage
$\rho$ as
\begin{equation}
	N_A \cong \rho \frac{O_S}{C_A}
		= 4 \rho \left(\frac{R}{r}\right)^2 \quad .
\label{eqNa}
\end{equation}
Note that the factor $2$ due to covering only one host particle and
the factor $\frac{1}{2}$ due to covering only half this particle
cancel each other out.

\subsection{Positioning}

After complete preparation of the first sphere, the second
uncovered sphere approaches along the $z$-axis in negative
direction, until it hits the covered hemisphere. In order to
determine the three contact points as easy as possible, the
coordinates of the $N_A$ adsorbates are transformed by projection on
the $x$-$y$-plane, $(x,y,z)\rightarrow(x,y)$, which is a sufficient
approximation for $r\ll R$.

\begin{figure}
	\includegraphics*{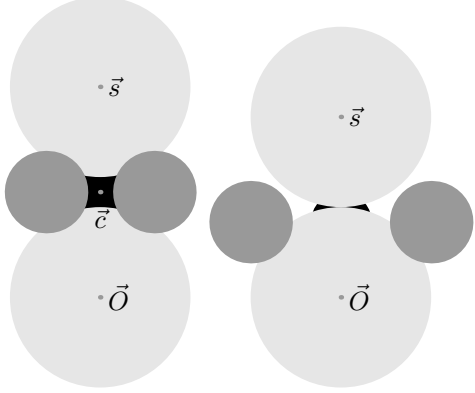}
\caption{\picleft If the circumcenter $\vec{c}$ of the contact
adsorbates is more than $D/2$ away from the surface of the covered
powder particle (located at the origin $\vec{O}$), the second powder
particle is placed in such a way that $\vec{c}$ bisects the distance
between these two powder particles. \picright If $\vec{c}$ is less
than $D/2$ away from the surface of the covered powder particle,
i.e., that the two powder particles will touch each other, the
second powder particle is placed in the direction of $\vec{c}$ so
that there is just the contact distance $D$ between the two powder
particles.
}
\label{figPosScheme}
\end{figure}

The roll-off mechanism described in \figur\ref{figPositioning} and
its caption yields the three contact points; consequently the
contact position of the second powder particle can be calculated,
which will be explained in the following. With the coordinates of
the centers of the three contact adsorbates the circumcenter
$\vec{c}$ of the triangle spanned by these three points can easily
be constructed. Note that the center of the covered powder particle
was defined to be located at the origin. So for symmetry reasons,
see \figur\ref{figPosScheme}, the center position $\vec{s}$ of the
second, uncovered particle can be computed as
\begin{equation}
	\vec{s} = \left\{
	\begin{array}{ll}
		2 \, \vec{c} \quad
		& \mbox{if} \quad |\vec{c}|\ge R+\frac{D}{2} \\[1ex]
		(2R+D) \, \frac{1}{|\vec{c}|} \, \vec{c} \quad
		& \mbox{if} \quad |\vec{c}|<R+\frac{D}{2}
	\end{array} \right. \quad .
\end{equation}
As the algorithm for finding the three-point contact does not check
for a direct contact between the two host particles, the artificial
definition of $\vec{s}$ for $|\vec{c}|<R+\frac{D}{2}$ must be
introduced. So we ensure a minimum distance of $D$ between the two
host particles. For this case of contacted powder particles the
adsorbates are of minor importance, since the direct powder-powder
interaction exceeds any powder-adsorbate interaction by some orders
of magnitude.

\section{Conversion of the parameters from
	Meyer's model}\label{appMeyer}

\begin{figure}
	\includegraphics*{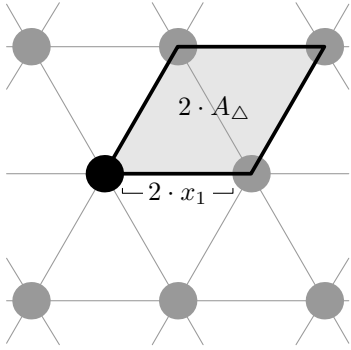}
\caption{Surface layout according to Meyer's model. The
nanoadsorbates (large dots) occupy the vertices of the equilaterally
triangulate surface of the powder particle (white background).
Distributing the surface area equally to the adsorbates, the area of
the highlighted cell belongs to each individual one.
}
\label{figMeyer}
\end{figure}

In order to be able to compare Meyer's model with our simulated
data, one has to convert the control parameter used in this study,
the relative surface coverage $\rho$, into the quantity $x_1$ used
by Meyer, cf.\ \figur\ref{figMeyer} and
\cite{zimmermann_et_al,diss_meyer}.

\gleichung{\ref{eqNa}} yields the total number of
adsorbates on one host particle, $\rho\mapsto N_A$. Although it is
not possible to cover the surface of a sphere globally by a
hexagonal lattice it is assumed that such a structure can be
established locally. According to Meyer's model, adsorbates are set
on the vertices of this lattice, so that a uniform pattern of
equilateral triangles evolves. To each adsorbate belongs a cell of
the area
\begin{equation}
	2\cdot A_{\triangle} = \frac{4\pi R^2}{N_A} \quad ,
\end{equation}
cf.\ \figur\ref{figMeyer}. With elementary geometrical considerations
one can calculate the side length of an equilateral triangle of the
area $A_{\triangle}$ as
\begin{equation}
	a = \frac{2}{\sqrt[4]{3}} \cdot \sqrt{A_{\triangle}} \quad .
\end{equation}
Consequently
\begin{equation}
	x_1 = \frac{a - 2r}{2} = \frac{a}{2} - r
\end{equation}
follows, which can be put in the formulas for $h(x_1)$ and
$F_{\mathrm{vdW}}(h)$, see \gleichungen{27} and (24) in
\cite{zimmermann_et_al}. Hereby a mapping
$\rho\mapsto F_{\mathrm{vdW}}$ gets feasible.

\section{Experiments}\label{appExperiments}

\subsection{Materials and methods}

Corn starch, Cerestar\textsuperscript{\textregistered} GL 03406, was
purchased from Cerestar Germany, Krefeld. The spherical particles
have diameters of $22\,\mathrm{\mu m}$. They are very cohesive. As
nanomaterials we used Aerosil\textsuperscript{\textregistered} 300
and Aerosil\textsuperscript{\textregistered} OX 50 both supplied by
DEGUSSA AG, Hanau, Germany. Their specific properties are summarized
in \tabelle\ref{tabSpecProp}.

\begin{table*}
\begin{tabular}{|l|c|c|c|}
	\hline
	Nanomaterial						&
	Diameter of the primary particles 			&
	Solid density 						&
	Specific surface 					\\
								&
	$[\mathrm{nm}]$						&
	$[\mathrm{g}/\mathrm{cm}^3]$				&
	$[\mathrm{m}^2/\mathrm{g}]$				\\
	\hline
	Aerosil\textsuperscript{\textregistered} 300		&
	$7$							&
	$2.2$							&
	$300\pm 30$						\\
	\hline
	Aerosil\textsuperscript{\textregistered} OX 50		&
	$40$							&
	$2.2$							&
	$50\pm 15$						\\
	\hline
\end{tabular}
\caption{Specific properties of the used nanomaterials.
}
\label{tabSpecProp}
\end{table*}

Both types of Aerosil\textsuperscript{\textregistered} are
hydrophilic. Due to their small sizes they highly agglomerate.

In our experiments we determined the tensile strengths of powder
mixtures consisting of corn starch and $0.2\%$ of a nanomaterial.
In order to prepare these mixtures first the nanomaterial was given
into a glass vial with a volume of $500\,\mathrm{ml}$. Then in a
second step $100\,\mathrm{g}$ of corn starch were added. The degree
of filling was given by $0.4$. The powders were mixed by means of a
Turbula mixer (Type T2C, \nummer{950353}, W. Bachofen AG, Basel,
Switzerland) at $42\,\mathrm{rpm}$. A comparison of the particle
sizes of the mixture components suggests that the mixing process
involves a grinding of the agglomerates formed by the nanomaterials.

\subsection{Measurement of tensile strengths}

The interparticle forces acting in the mixtures consisting of corn
starch and $0.2\%$ of a nanomaterial were measured by means of
tensile strength tester developed by Schweiger
\cite{diss_schweiger,schweiger_zimmermann} and modified by Anstett
\cite{anstett}, \figur\ref{figTensileTester}.

The powder to be tested is filled into the sample container having
a volume of $4.62\,\mathrm{cm}^3$ (diameter: $3.43\,\mathrm{cm}$,
height: $0.50\,\mathrm{cm}$). By means of a sieve with a sieve size
of $315\,\mathrm{\mu m}$ it is sieved directly into the sample
container. In order to achieve a homogenous filling the filled
container is gently tapped onto a soft pad. By scraping off the
surplus testing material by means of a slide a perfectly flat
surface of the powder bed can be obtained. This step has to be
performed very carefully in order to avoid even the slightest
compression of the powder.

The sample container prepared in this way is placed on a small table
which can be moved up and down by means of the linear positioner
M-410-21 (Physik Instrumente GmbH\&Co, Waldhorn). By coarse
adjustment screws the whole unit consisting of a table as well as a
linear positioner is moved upwards so that only a very small gap of
about $1$ to $2\,\mathrm{mm}$ remains between the powder surface and
the measuring plate. From this position on the further movement of
the table by the linear positioner is controlled by the controller
C-832.00 (Physik Instrumente GmbH\&Co, Waldhorn) which runs under
the software package Pro Move v2.20. The steps performed by the
linear positioner can be simultaneously seen on the screen of the
PC. With a vertical speed of $3.4\,\mathrm{\mu m}/\mathrm{s}$ the
table is moved upwards until the measuring plate comes into contact
with the powder surface. The movement is stopped as soon as a
defined negative weight of the measuring plate is measured.

The cylindrical measuring plate has a lower surface of
$0.385\,\mathrm{cm}^2$. By means of a frame made from Aluminum it is
fixed on a load cell ZER 10 (Wipotec GmbH, Kaiserlautern, Germany).
The measuring range of this load cell covers a range of
$2\,\mathrm{g}$ with a resolution of $0.1\,\mathrm{mg}$. The force
acting on the load cell is registered by the software
Wipotec-Terminal v3.78 at a scan rate of three measurements per
second. The signals of the load cell are mass calibrated. In order
to obtain the corresponding forces the signals have to be
multiplied by the gravitational constant $g$.

Before each measurement a solution of $5\%\,(w/w)$ white
petrolatum in petrolether is sprayed onto the lower side of the
measuring plate. After evaporation of the solvent the measuring
plate is coated with a thin layer of petrolatum as a sticking agent
to ensure that the powder particles adhere on its surface.

\begin{figure}
	\includegraphics*{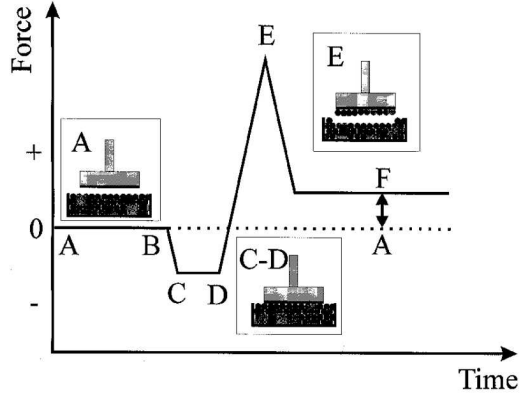}
\caption{Time course of the force during a measuring cycle.
}
\label{figTimeCourse}
\end{figure}

After a pause of $10\,\mathrm{s}$ the table with the sample
container is moved downwards at a speed of
$1.7\,\mathrm{m}/\mathrm{s}$. Due to the interaction between the
layer of powder particles fixed at the lower side of the measuring
plate and the particles remaining in the adjacent powder layer in
the sample container the load cell measures a force. This force
increases with continuing downwards movement of the sample. It
reaches its maximum when the two powder layers separate. The force
measured by the load cell drops to that force corresponding to the
weight of the powder layer adhering at the lower surface of the
measuring plate, \figur\ref{figTimeCourse}.

In order to calculate the tensile strength $\sigma$ the force $F_F$
measured at point F is subtracted from the maximum force $F_E$
measured at point E. This value is then divided by the lower
surface $A_M$ of the measuring plate:
\begin{equation}
	\sigma = \frac{F_E - F_F}{A_M} \quad .
\end{equation}


\end{document}